\documentclass[aps,prl,twocolumn,showpacs,superscriptaddress,floatfix]{revtex4-1}
\usepackage{bm}
\usepackage{amssymb}
\usepackage{graphicx}
\usepackage{amsmath}
\usepackage{dcolumn}
\usepackage{bm}
\usepackage{color}
\usepackage[german,english]{babel}

\newcommand{\bk}{\mathbf{k}}
\newcommand{\bq}{\mathbf{q}}

\renewcommand{\Re}{{\rm Re}\,}
\renewcommand{\Im}{{\rm Im}\,}
\newcommand{\beq}{\begin{eqnarray}}
\newcommand{\eeq}{\end{eqnarray}}
\newcommand{\beqa}{\begin{equation}}
\newcommand{\eeqa}{\end{equation}}

\begin{document}

\title{Enhanced Screening in Chemically Functionalized Graphene}
\author{Shengjun Yuan}
\email{s.yuan@science.ru.nl}
\affiliation{Radboud University of Nijmegen, Institute for
Molecules and Materials, Heijendaalseweg 135, 6525 AJ Nijmegen,
The Netherlands}
\author{T. O. Wehling}
\email{wehling@itp.uni-bremen.de}
\affiliation{Institut f{\"u}r Theoretische Physik, Universit{\"a}t Bremen, Otto-Hahn-Allee 1, 28359 Bremen, Germany}
\affiliation{Bremen Center for Computational Materials Science, Universit{\"a}t Bremen, Am Fallturm 1a, 28359 Bremen, Germany}
\author{A. I. Lichtenstein}
\affiliation{Institut f{\"u}r Theoretische Physik,
Universit{\"a}t Hamburg, Jungiusstra{\ss}e 9, D-20355 Hamburg,
Germany}
\author{M. I. Katsnelson}
\affiliation{Radboud University of Nijmegen, Institute for
Molecules and Materials, Heijendaalseweg 135, 6525 AJ Nijmegen,
The Netherlands}
\date{\today}

\begin{abstract}
Resonant scatterers such as hydrogen adatoms can strongly enhance the low energy density of states in graphene. Here, we study the impact of these impurities on the electronic screening. We find a two-faced behavior: Kubo formula calculations reveal an increased dielectric function $\varepsilon$ upon creation of midgap states but no metallic divergence of the static $\varepsilon$ at small momentum transfer $q\to 0$.  This bad metal behavior manifests also in the dynamic polarization function and can be directly measured by means of electron energy loss spectroscopy. A new length scale $l_c$ beyond which screening is suppressed emerges, which we identify with the Anderson localization length.
\end{abstract}

\maketitle
Electronic screening presents a central problem in the physics of graphene and strongly affects electron transport as well as effects of electron-electron
interactions in this material. First,
external perturbations like charged impurities in the graphene substrate can
be screened by the graphene electrons and the resulting scattering rate
becomes inversely proportional to the squared effective dielectric constant $\varepsilon
$ of graphene \cite{DasSarma_RMP11,Katsnelson_book}. 
Second, electron-electron interactions are known to renormalize the
charge carrier velocity in graphene near the Dirac point \cite{Guinea_94,Elias_NaturePhys10} and an excitonic
instability can be realized depending on $\varepsilon$ \cite{Drut_Lahnde_PRL2009}. 

In reality, graphene samples are subject to disorder, which can alter $\varepsilon $ significantly. Single impurities can, for instance, lead to high energy plasmon poles of $\varepsilon$ \cite{Balatsky_Plasmons_PRB10}. Most notably, impurities such as hydrogen adatoms can
lead to pronounced resonances in the density of states near the Dirac point
and related impurity bands can largely increase the low energy density of
states (DOS) for impurity concentrations on the order of a few percent or
less \cite{Pereira_PRL2006,Peres_PRB06,Wehling_PRB07,Wehling_CPL09,WK10,Katsnelson_book}. At high impurity coverages, however, graphene can turn into an insulating
material such as graphane \cite{Elias_Science09} or fluorographene \cite{Nair_small2010} with band gaps on the order of
some eV \cite{Klintenberg_11} and thus vanishing DOS at low energies. Regarding the screening of
electric fields in graphene it is unclear whether chemical
functionalization with species like hydrogen metallizes graphene or rather turns it into an insulator. The answer to question is complicated by the special physics of Anderson localization in this material: The chiral symmetry has been shown to suppress Anderson localization in charge neutral graphene even in presence of strong local impurities \cite{Mirlin_RMP08,Ostrovsky_PRL2010}. Away from the neutrality point, however, hydrogen adatoms can lead to Anderson localization \cite{Chang_PRB2010}.

In this letter, we study electronic screening in graphene with resonant scatterers by means of numerically exact Kubo formula calculations \cite{YRK10,YRK11}. The dielectric function $\varepsilon (q,\omega)$
and the dynamic polarization function $\Pi (q,\omega )$ are discussed as function of momentum
transfer $q$ and frequency $\omega$. As the central result, we find that resonant impurities
turn graphene into a bad metal: Screening in graphene becomes indeed enhanced by chemical functionalization, as the static dielectric
function $\varepsilon (q,\omega =0)\equiv\varepsilon(q)$ exceeds the pristine graphene value of $\varepsilon _{G}(q)\approx 4.93$. Below a critical length scale $l_{c}$, we find metallic behavior: $\varepsilon (q)\sim q^{-1}$ for $q>q_{c}=1/l_c$. However, $\varepsilon (q)$ exhibits a maximum at $q\sim
q_{c}$ and decreases with $q$ for $q<q_{c}$. Chemically functionalized graphene thus
provides metallic screening only at length scales below $l_{c}$. 
We further find bad metal characteristics in the dynamic polarization function and in $-\Im1/\varepsilon(q,\omega)$ which can be measured by means of electron energy loss spectroscopy (EELS). By comparing to Green function calculations which neglect vertex corrections, we show that the bad metal behavior is not due to redistribution of single particle spectral weight but due to quantum interference processes. $l_c$ is identified with the Anderson localization length.

To describe graphene functionalized with chemical species like hydrogen or organic groups, we consider the Hamiltonian $H=H_{gr}+H_{imp}$. $H_{gr}=-t%
\sum_{<i,j>}(a_{i}^{\dagger }b_{j}+\mathrm{h.c})$ is nearest-neighbor tight-binding Hamiltonian of graphene, where $a_{i}^{\dagger }$ (%
$b_{i}$) creates (annihilates) an electron on sublattice A (B) and $t=2.7eV$ is the nearest neighbor hopping parameter.
The adsorbates are taken into account through the Hamiltonian $H_{imp}=\epsilon_{d}\sum_{i}d_{i}^{\dagger}d_{i}+V\sum_{i}\left( d_{i}^{\dagger}c_{i}+\mathrm{h.c}\right)$, where the parameters $V=2t$, $%
\epsilon _{d}=-t/16$ for hydrogen have been determined by density functional
theory (DFT) calculations \cite{WK10}. These impurities lead to resonances in the spectrum of graphene and to the formation of an impurity band, which is energetically very close to the Dirac point \cite{Pereira_PRL2006,Peres_PRB06,Wehling_PRB07,Pereira2008,Wehling_CPL09,WK10,YRK10}. Recently these impurity states have been
observed experimentally by means of scanning tunneling spectroscopy (STM) \cite{Ugeda2010} and photoemission spectroscopy \cite{Haberer2011}.



The dynamical polarization function can be obtained from the Kubo formula \cite{K57}
as%
\begin{equation}
\Pi \left( \mathbf{q},\omega \right) =\frac{i}{A}\int_{0}^{\infty }d\tau
e^{i\omega \tau }\left\langle \left[ \rho \left( \mathbf{q},\tau \right)
,\rho \left( -\mathbf{q},0\right) \right] \right\rangle ,  \label{Eq:Kubo}
\end{equation}%
where $A$ denotes the area of the unit cell, $\rho \left( \mathbf{q}\right)
=\sum_{i}c_{i}^{\dagger }c_{i}\exp \left( i\mathbf{q\cdot r}_{i}\right) $ is
the density operator, and the average is taken over the canonical ensemble.
For the case of a single-particle Hamiltonian Eq.~(\ref{Eq:Kubo}) can be
calculated by means of the Chebyshev polynomial method \cite{YRK10,YRK11}. The screening of electric fields is
determined by the dielectric function $\mathbf{\varepsilon }\left( \mathbf{q}%
,\omega \right) $. In the random phase approximation (RPA) it reads $\mathbf{\varepsilon }\left( \mathbf{q},\omega \right) =\mathbf{1}-V\left( q\right)
\Pi \left( \mathbf{q},\omega \right) $, where $V\left( q\right) =2\pi
e^{2}/ q$ is the Fourier component of the Coulomb interaction in two
dimensions. 

\begin{figure}[t]
\begin{center}
\mbox{
\includegraphics[width=0.8\columnwidth]{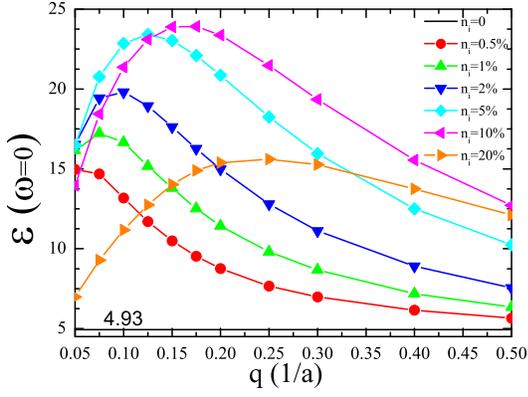}
}
\end{center}
\caption{(Color online) Static dielectric function of graphene with
different concentrations of impurities $n_i$ as function of wave vector $q$ at fixed
chemical potential $\mu=0$. The maxima in $\varepsilon(q,\omega=0)$ indicate bad metal behavior. The wave vector $\mathbf{q}$ is in the $\Gamma -K$
direction. Periodic boundary conditions are used in a sample containing $4096\times 4096$ carbon atoms.}
\label{staticdielectricfunction}
\end{figure}

We start with analyzing the static dielectric function (Fig. \ref{staticdielectricfunction}) obtained from our Kubo formula calculations. For pristine graphene, the RPA static dielectric function $\varepsilon_G(q)=4.93$ is a constant \cite{WSSG06}, which is also found in our calculations. In the presence of hydrogen impurities, one may expect that the screening is highly enhanced due to the
emergence of midgap states in the vicinity of the neutrality point and the correspondingly increased density of states around the Fermi level. Indeed, this expectation holds for impurity concentrations up to $n_i\lesssim 20\%$ and the $\bq$-vectors ($q\geq 0.05/a$, where $a$ is carbon-carbon distance) accessible in our simulations \footnote{Due to limitations of present computer power, it is numerically too expensive to get reliable values of the polarization function for wave vectors $q<0.05/a$}. However, the shape of the $\varepsilon(q)$-curves does not show purely metallic behavior (i.e. $\varepsilon(q)\sim q^{-1}$), as might be expected from the non-zero density of states at the Fermi level: We find $\varepsilon (q)\sim q^{-1}$ but only for $q>q_{c}$ with impurity concentration dependent $q_c$. $\varepsilon (q)$ exhibits a maximum at $q\sim q_{c}$ and decreases for $q<q_{c}$. Electronic screening in graphene is thus enhanced by hydrogen functionalization but fully metallic screening is found only at length scales below $l_{c}\sim 1/q_{c}$.

\begin{figure}[t]
\begin{center}
\mbox{
\includegraphics[width=0.5\columnwidth]{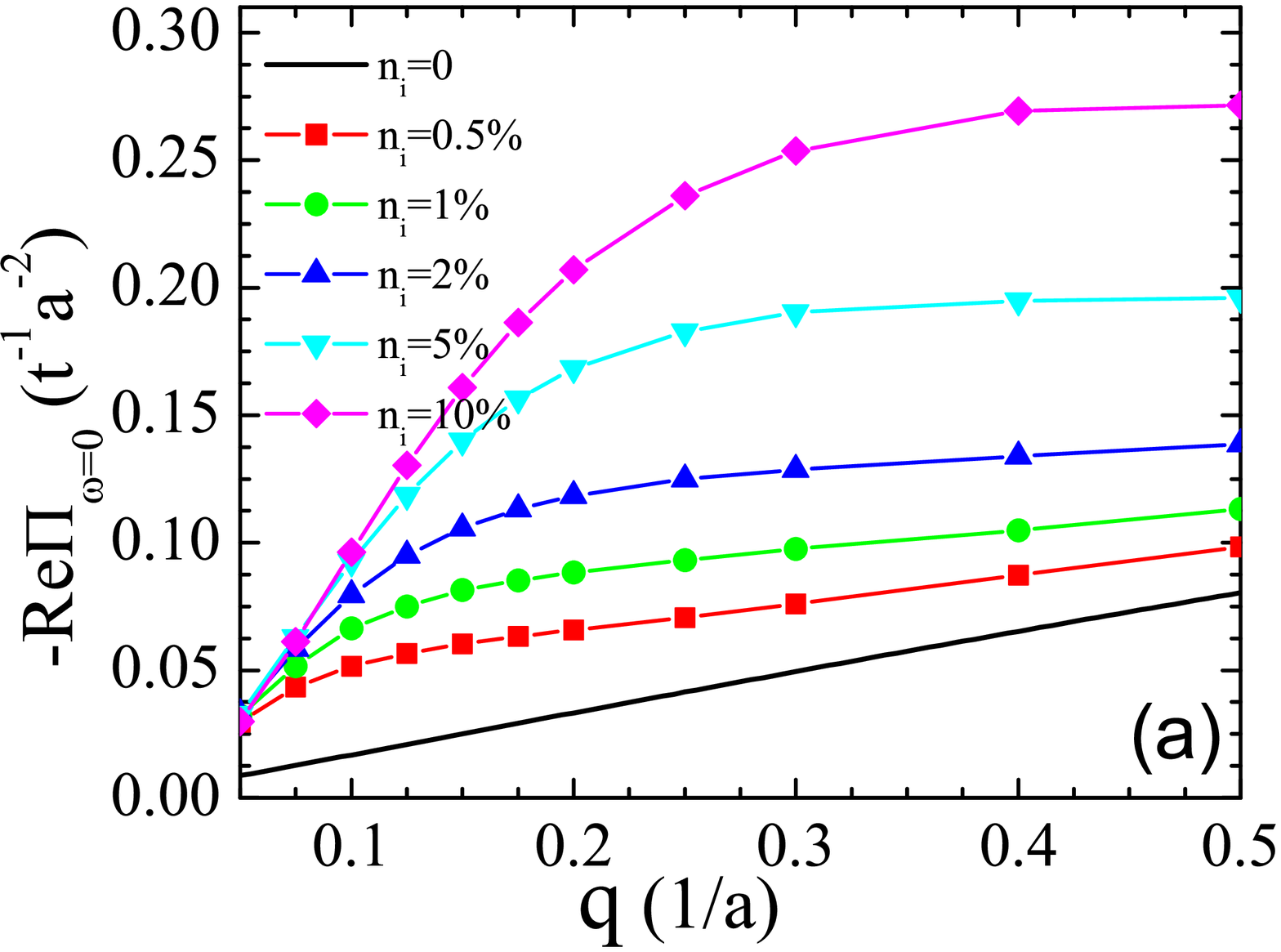}
\includegraphics[width=0.5\columnwidth]{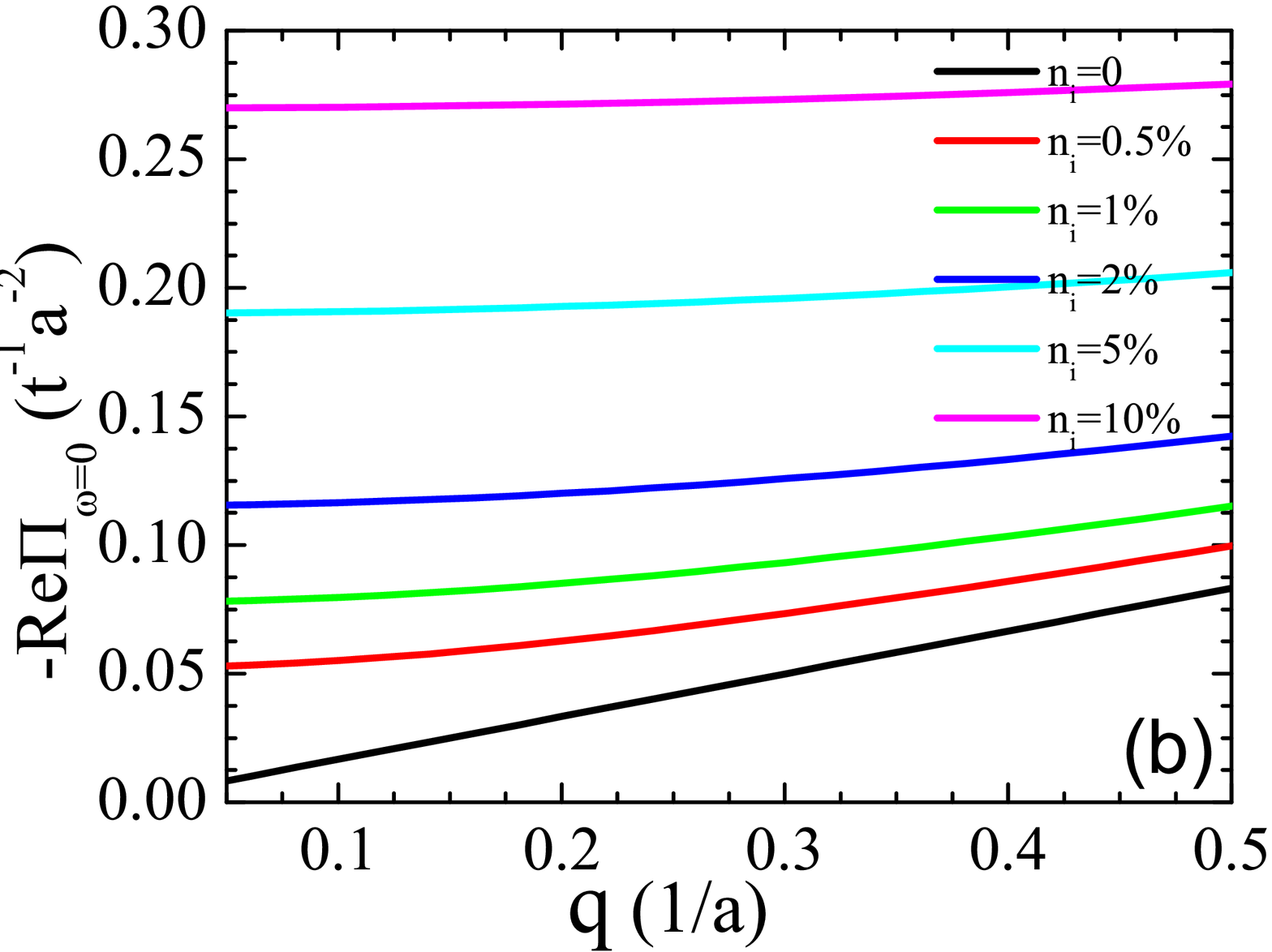}
} 
\mbox{
\includegraphics[width=0.5\columnwidth]{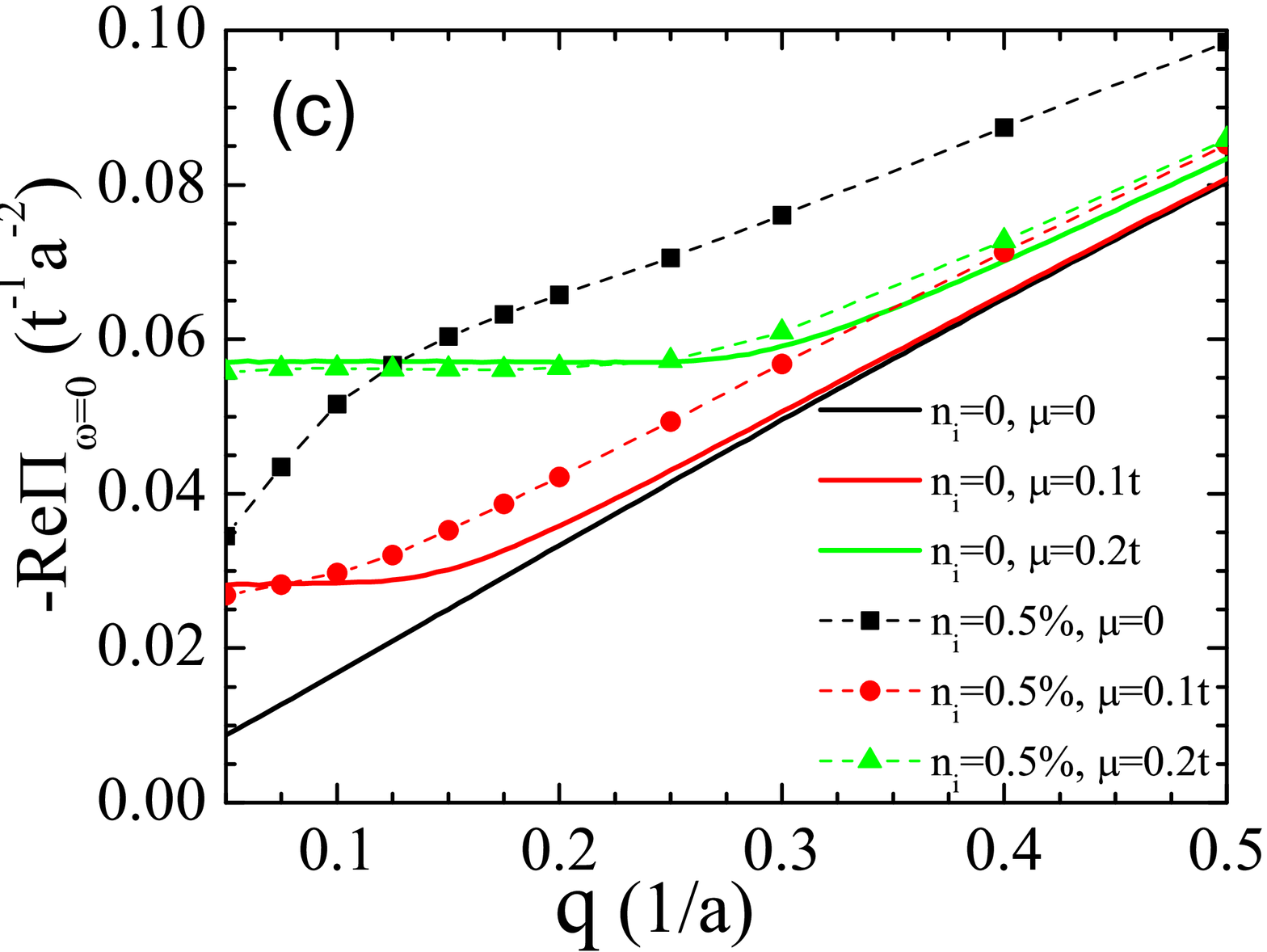}
\includegraphics[width=0.5\columnwidth]{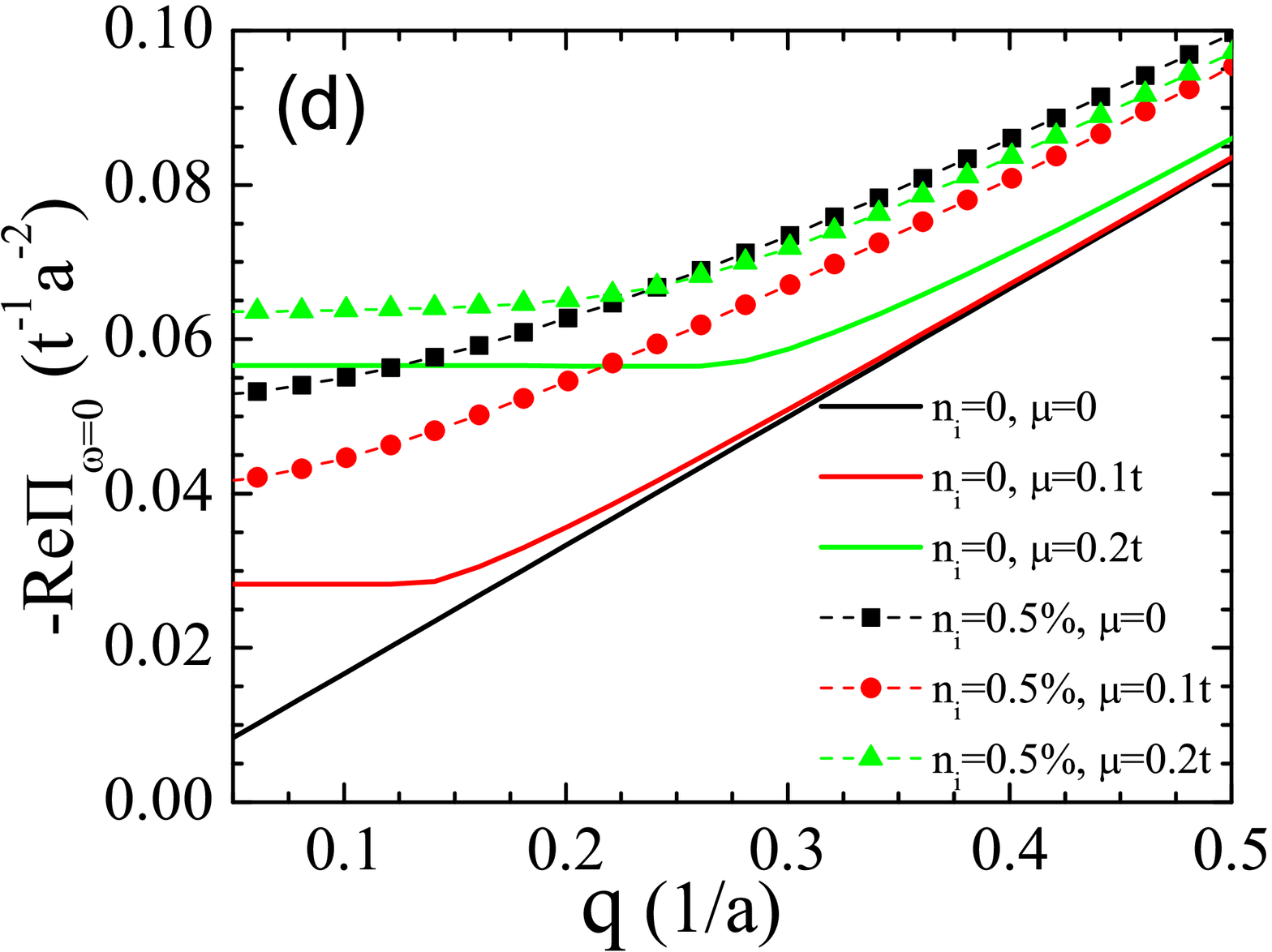}
}
\end{center}
\caption{(Color online) (a,b) Static polarization function of graphene
with different concentrations of impurities $n_i$ as a function of wave vector $q$.
The chemical potential is fixed as $\mu=0$. (c,d) $q$-dependent static polarization
function of graphene at different chemical potentials $\mu=0$, $0.1t$, and $0.2t$. 
Results for graphene without impurities (solid lines) are compared to those with an impurity concentration of $n_{i}=0.5\%$ (dashed lines). 
(a,c) Results from the numerically exact Kubo formula calculations. (b,d)
Results obtained by neglecting vertex corrections (Eq. (\ref{eq:Gf_Pi})).}
\label{sp_miu}
\end{figure}

To understand the origin of this effect, we analyze the static polarization function ($\Pi(q,\omega=0)\equiv \Pi(q)$, Fig. \ref{sp_miu}) obtained from the Kubo formula calculations and compare to the results obtained from evaluating a bubble diagram,
\begin{equation}
\Pi (\mathbf{q},i\omega _{n})=\frac{1}{A\beta }\sum_{\mathbf{k},n^{\prime }}%
\mathrm{Tr}\,G(\mathbf{k},i\omega _{n^{\prime }})G(\mathbf{k}+\mathbf{q}%
,i\omega _{n^{\prime }}+i\omega _{n}).  \label{eq:Gf_Pi}
\end{equation}%
Here, $G(\bk,i\omega_n)=(i\omega_n+\mu-v_{\rm F}(\sigma\cdot\mathbf{k})-\Sigma(i\omega_n))^{-1}$ is the Green function of graphene with Fermi velocity  $v_{\mathrm{F}}=3at/2$ in presence of strong local impurities, which are accounted for by the self-energy $\Sigma(i\omega_n)=n_i T(i\omega_n)$. In the limit of strong impurity potentials, the $T$-matrix entering the self-energy reads $T(i\omega_n)=-1/G^0_{\rm loc}(i\omega_n)$, where $G^0_{\rm loc}(i\omega_n)$ is the local Green function of pristine graphene. This formalism accounts for spectral weight redistribution on a single particle level. The only approximation of Eq. (\ref{eq:Gf_Pi}) is that it neglects vertex corrections which describe quantum interference effects like Anderson localization.

While both approaches yield the similar $\Pi(q)$ for larger $q$, we find that Eq. (\ref{eq:Gf_Pi}) yields constant non-zero $\Pi(q)$ for $q\to 0$ for graphene with hydrogen impurities, which corresponds to $\varepsilon\left( q=0\right) \rightarrow \infty $. This is in contrast to finite and decreasing $\varepsilon \left( q\right)$ and  decreasing $\Pi(q)$ for $q\to 0$ as found by the Kubo formula simulations. The deviation of the bubble-diagram from the numerically exact Kubo formula results is strongest if the chemical potential ($\mu\approx 0$) is within the midgap impurity band generated by the hydrogen adatoms (Fig. \ref{sp_miu}c and d). Thus, the bad metal behavior of $\varepsilon(q)$ for $q<q_c$ must be due to a quantum interference effect involving the hydrogen induced midgap states. For doped graphene ($\mu \neq 0$), the presence of the hydrogen impurities
becomes less important. 


\begin{figure}
\includegraphics[width=0.5\linewidth]{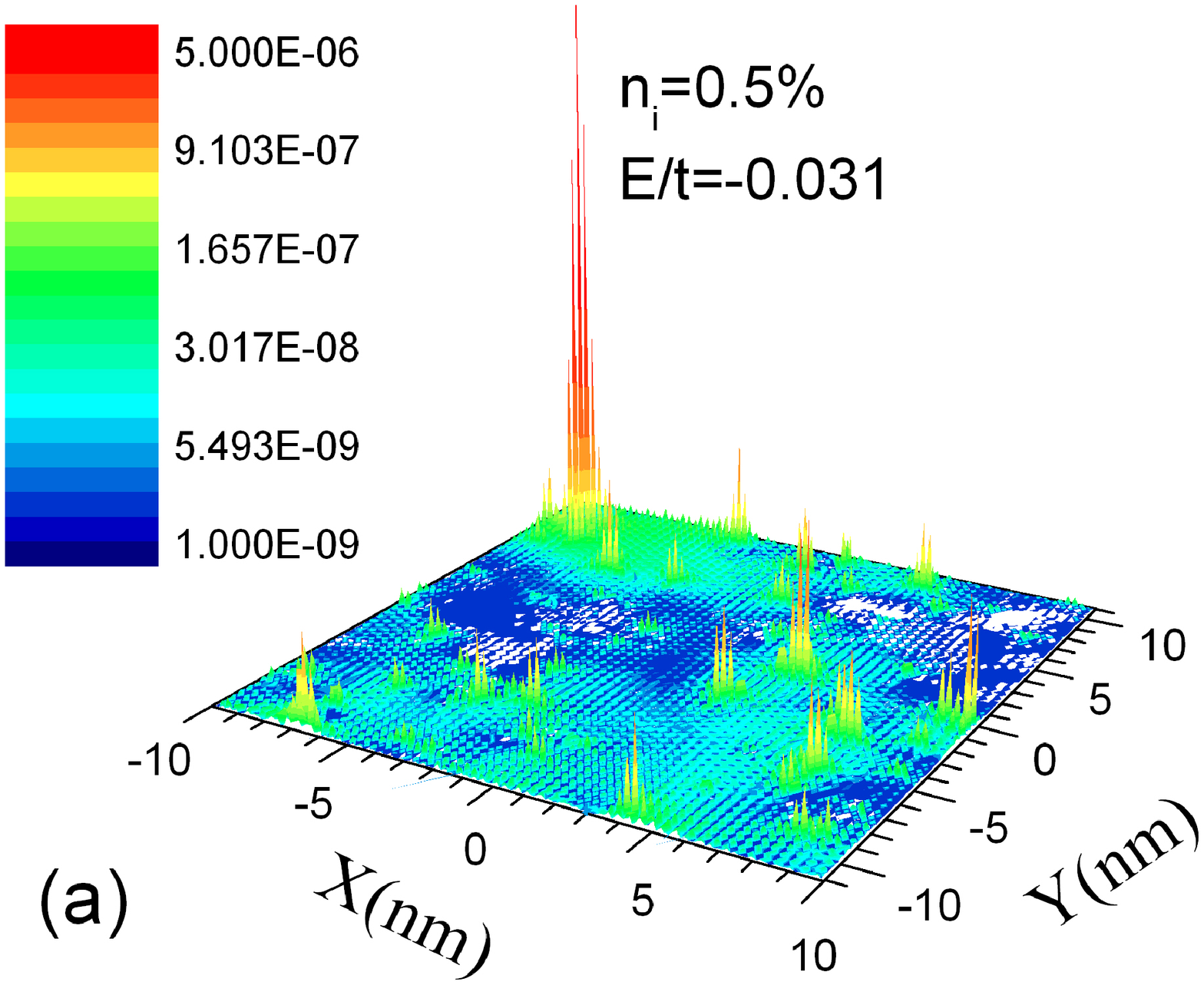}\includegraphics[width=0.5\linewidth]{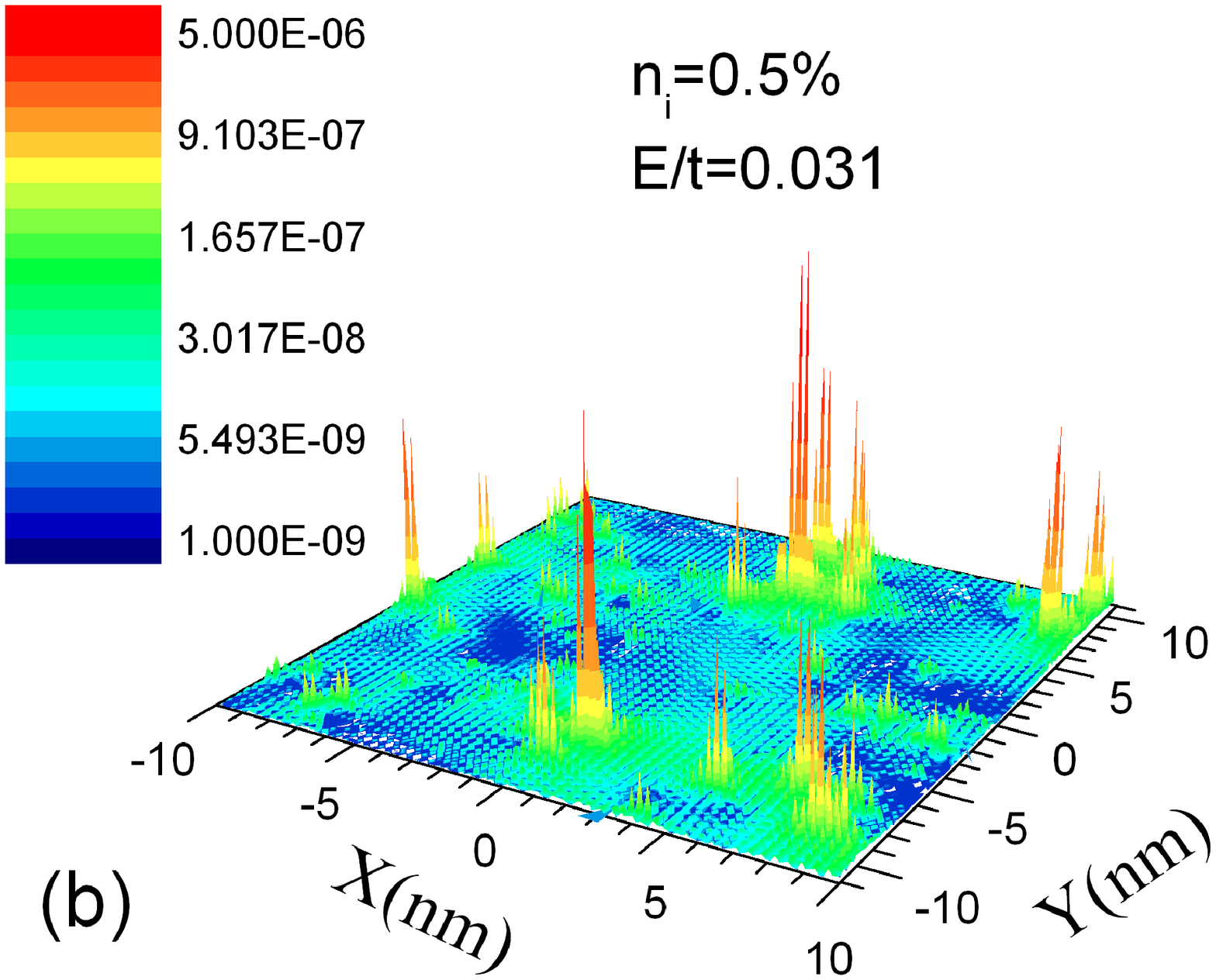}

\includegraphics[width=0.5\columnwidth]{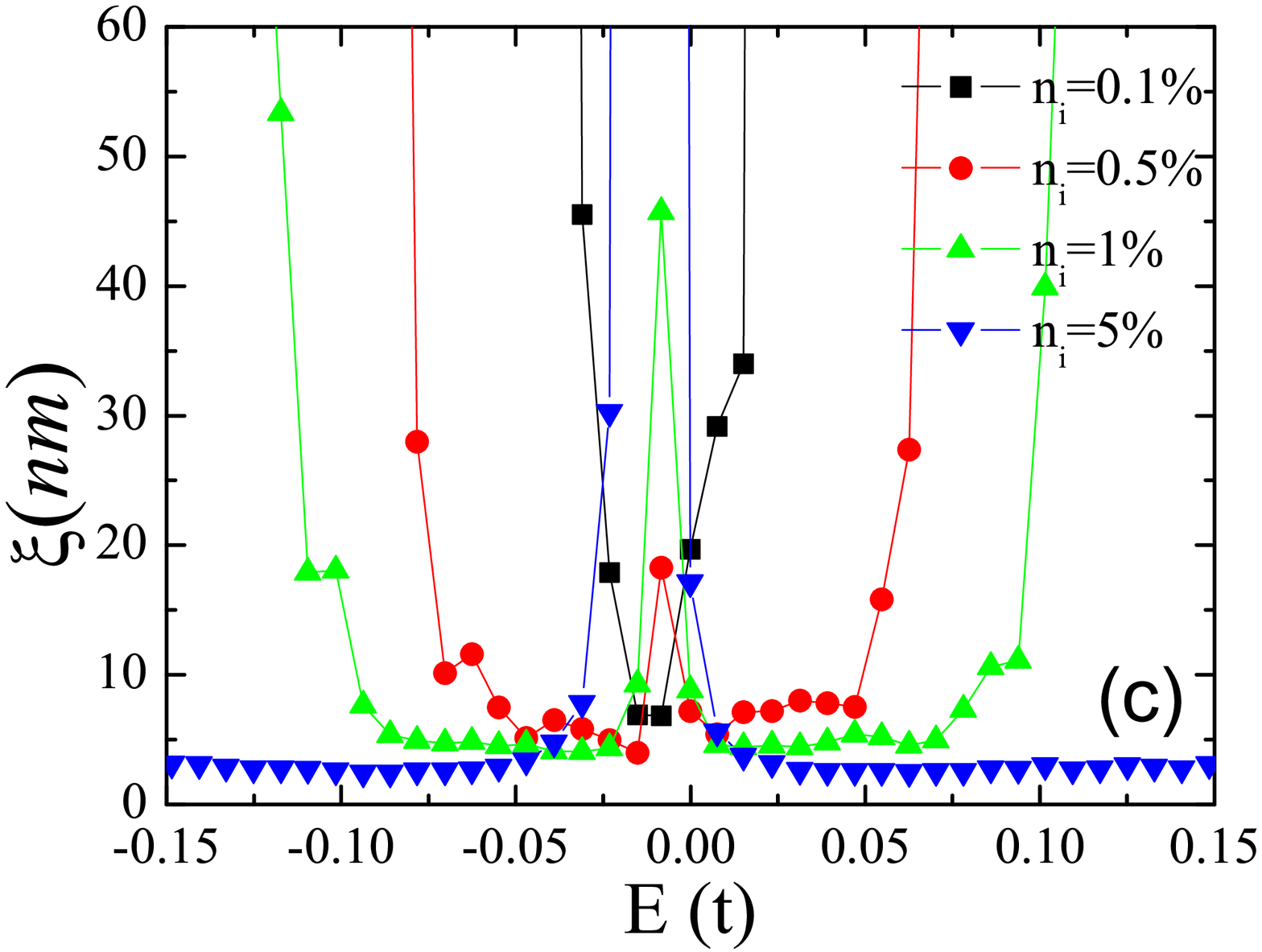}\includegraphics[width=0.5\columnwidth]{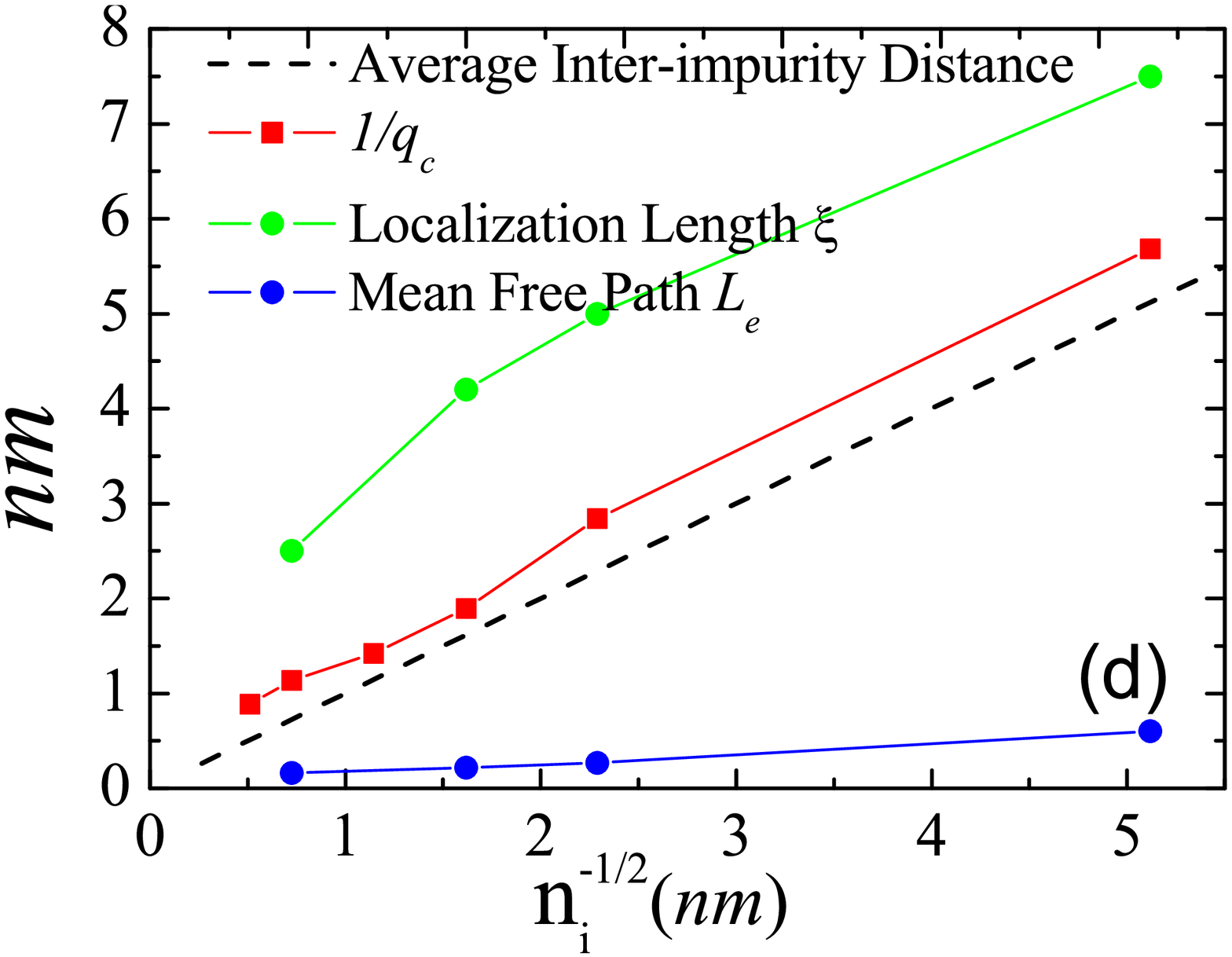}
\caption{(Color online) (a,b) Quasi eigenstates at energies $E=-0.031t$ (a) and $+0.031t$ (b) for graphene with hydrogen adatoms at concentration $n_i=0.5\%$.
(c) Energy dependent localization lengths in graphene with different concentrations $n_i$ of hydrogen adatoms.
(d) Critical screening length $l_c=1/q_c$, Anderson localization length $\xi$ and mean free path $L_E$ as function of average inter-impurity distance $1/\sqrt{n_i}$. $l_c$ is evaluated for $\mu=0$, whereas $\xi$ and $L_E$ are evaluated for $\mu$ inside the tails of the impurity band.}%
\label{fig:localization_lengths}%
\end{figure}

As Fig. \ref{fig:localization_lengths} (a,b) illustrates, the eigenstates in the energy range of the impurity band do not extend through the entire system but are localized in distinct regions of the sample. The spots where the quasi-eigenstates are localized depend strongly on energy, which is typical in the regime of Anderson localization \cite{Shklovskii_Efros}. To corroborate Anderson localization as physical origin of the suppressed long wavelength screening near the neutrality point quantitatively, we calculated electronic conductivities $\sigma$, diffusion coefficients $D(t,E)$ and the elastic mean free paths $L_e$ for undoped graphene ($\mu = 0$) with different concentrations of hydrogen impurities (see \cite{WK10,YRK10} for technical details). This allowed us to estimate the Anderson localization lengths $\xi=L_e\exp(\pi h \sigma/2e^2)$ \cite{Lee_RMP85, Mirlin_RMP08, Roche_PRL11}. Inside the energy range of
the midgap impurity states we find an interesting energy dependence of the localization lengths (Fig. \ref{fig:localization_lengths}c):
There is central peak showing comparably large localization lengths which paradoxially increase with impurity concentration. This peak is surrounded by plateaus, where the localization length is almost independent of the energy. In the energy range of these plateaus, the localization lengths decrease with increasing impurity concentration as one expects.

The comparison of the critical length scale for dielectric screening $l_c=1/q_c$ to the Anderson localization length $\xi$ inside the plateau region (Fig. \ref{fig:localization_lengths}d) reveals a very similar dependence of $l_c$ and $\xi$ on the impurity concentration. It is thus very likely that the suppressed screening at long wavelengths is due to Anderson localization in the midgap impurity band. The critical length scale $l_c$ and the Anderson localization length correspond almost exactly to the average inter-impurity distance. It appears that this is the only natural length scale induced by the impurities. 

The increase of localization lengths with the impurity concentration in the center of the impurity band does not manifest in any of the dielectric screening properties investigated, here. It is plausible that the high DOS in the center of the impurity band leads to shorter average hopping distances between localized  impurity states than in the tails of the impurity bands and might explain the increase of the conductivities / localization lengths with impurity concentration in that energy range. This mechanism is qualitatively similar to variable range hopping \cite{Shklovskii_Efros} with one difference. The latter requires some inelastic processes, usually electron-phonon scattering. Here, an energy uncertainty is provided by a general finite lifetime broadening: The semiclassical conductivities $\sigma$ correspond by definition \cite{Lee_RMP85, Mirlin_RMP08, Roche_PRL11} to the maximum conductivities obtained at finite simulation times and thus include an intrinsic lifetime broadening.

\begin{figure}[t]
\begin{center}
\includegraphics[width=0.48\columnwidth]{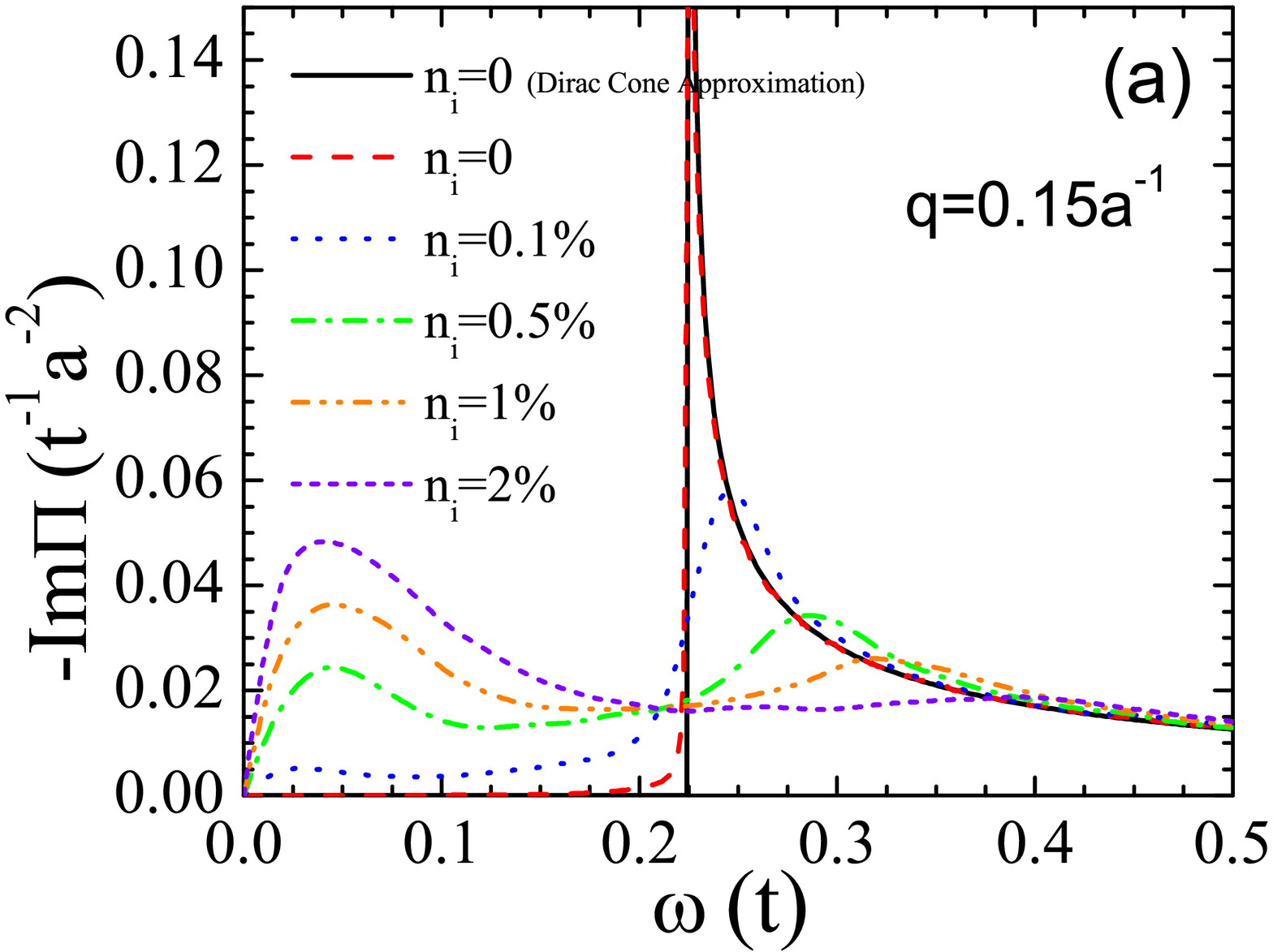}\hfill
\includegraphics[width=0.48\columnwidth]{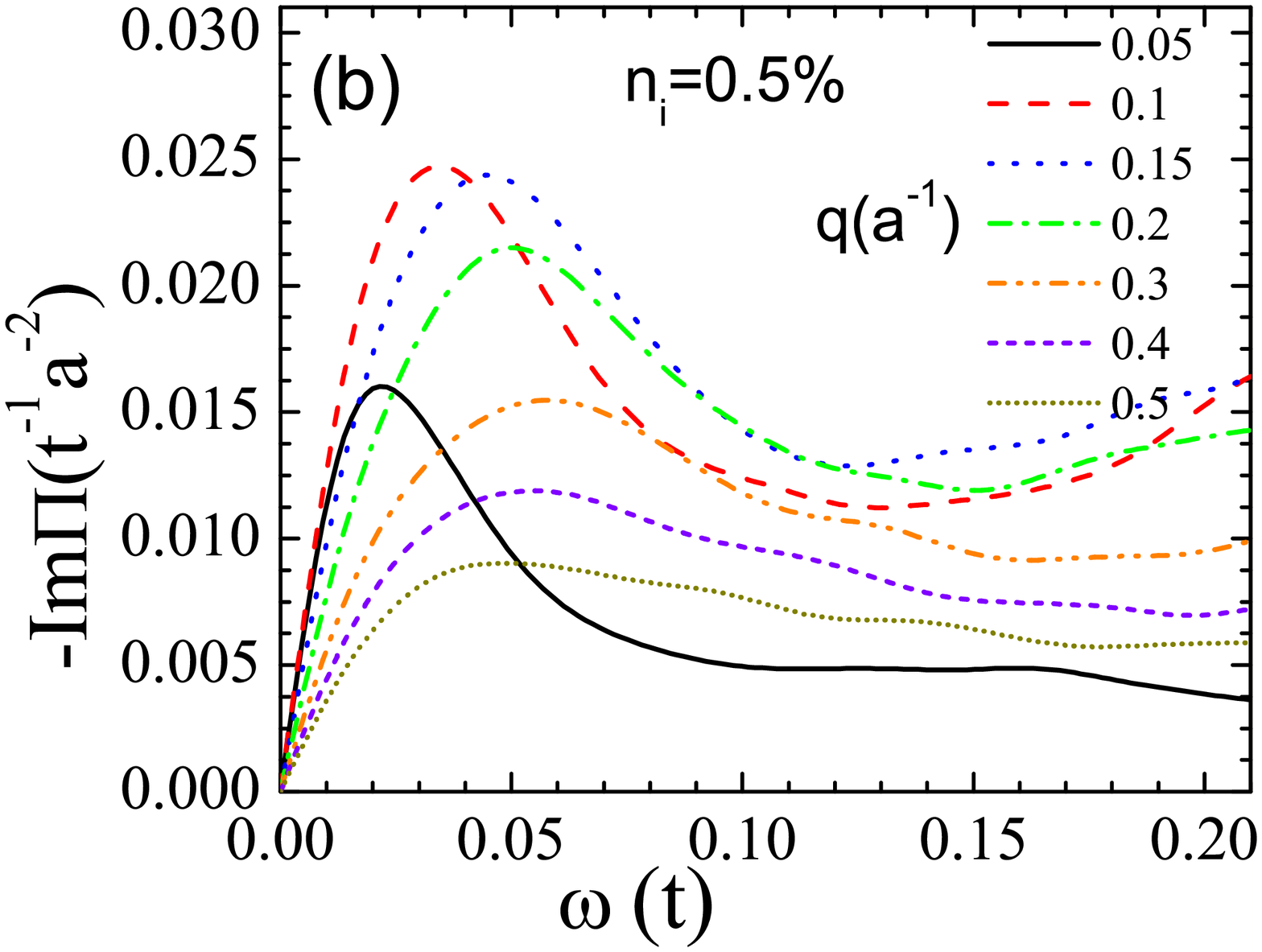}

\includegraphics[width=0.48\columnwidth]{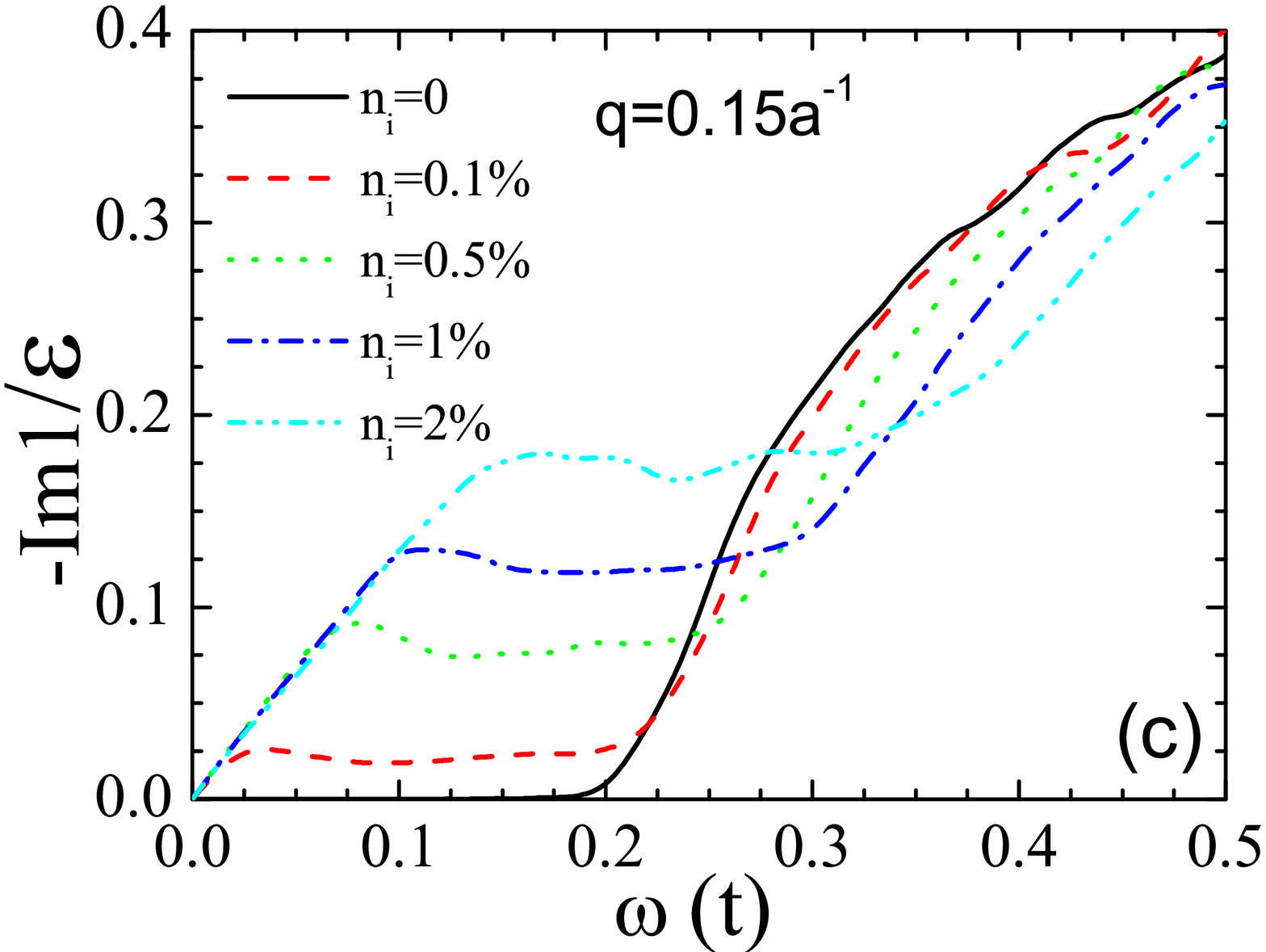}\hfill
\includegraphics[width=0.48\columnwidth]{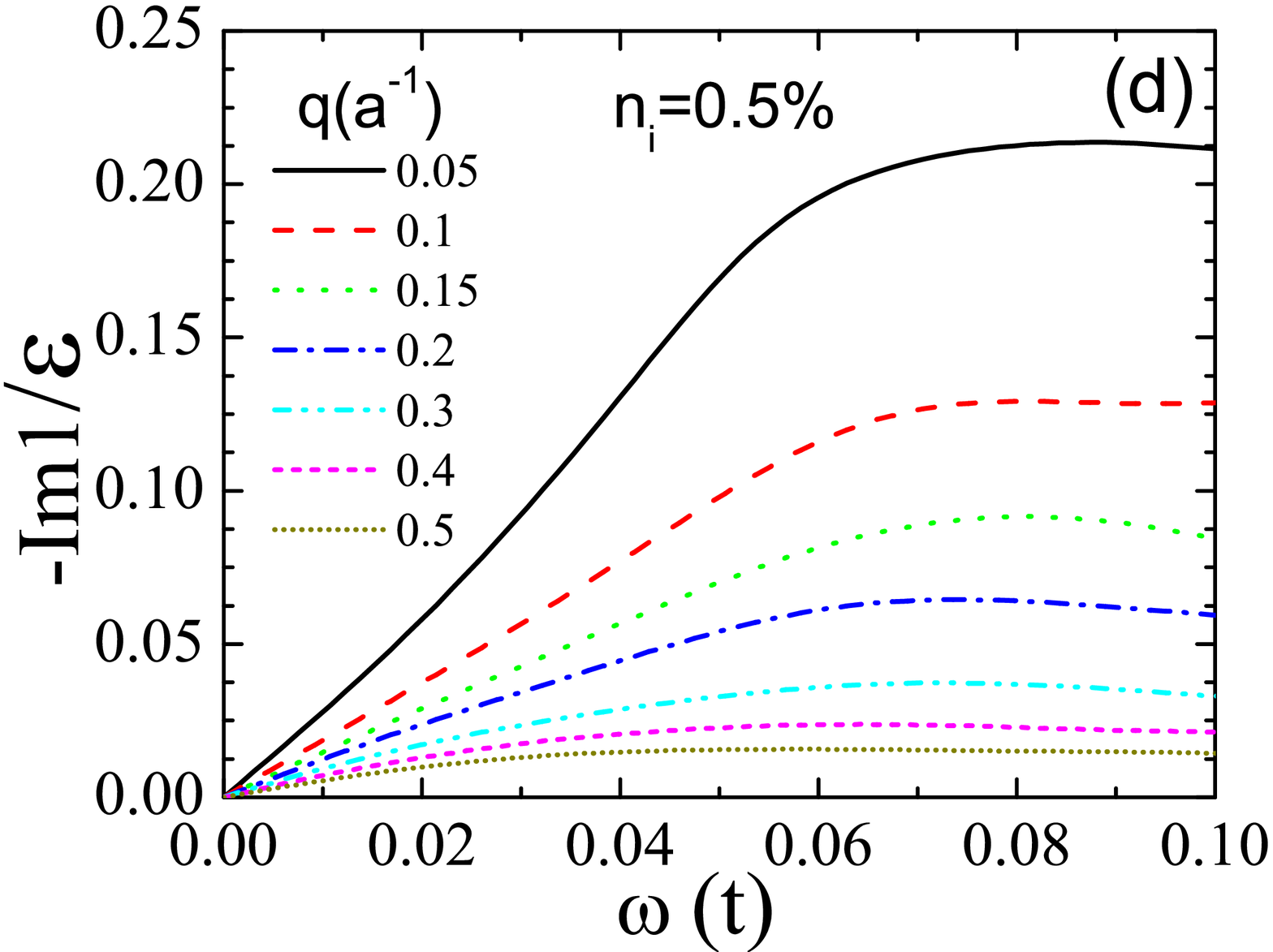}


\end{center}
\caption{(Color online) Dynamical polarization function of graphene at chemical potential ($\mu=0$) with (a)
different concentrations of impurities and the same wave vector $q=0.15a^{-1}$ as well as (b) with
different wave vectors for the same concentration of impurities. (c) Electron energy loss function $-\Im(1/\varepsilon)$ of graphene with adsorbed hydrogen impurities at concentrations between $n_i=0$ and $n_i=2\%$ with fixed $q=0.15a^{-1}$. There is impurity induced EELS intensity at energies $\omega<v_F q$, which is energetically forbidden in pristine graphene. (d) $-\Im(1/\varepsilon)$ in the low energy region for different momentum transfers $q$ at fixed impurity concentration $n_i=0.5\%$.}
\label{dpim_miu}
\end{figure}

The screening of static electric fields is determined by the generation of virtual electron hole pairs and, thus, involves processes from different energies. Being the real part of a retarded correlation function, $\Re\,\Pi (\mathbf{q},\omega=0)$ can include information about virtual electron hole pairs at arbitrary excitation energies. Here, finite-energy particle hole pairs contribute to $\Re\,\Pi (\mathbf{q},\omega=0)$ due to finite $\mathbf{q}$, two C-$p_z$ bands being present in graphene and due to the disorder. Therefore, it is understandable why transport localization lengths evaluated at energies different from the chemical potential in the screening calculations display the same trend as $l_c$ in Fig. \ref{fig:localization_lengths}. $\mathrm{Im}~\Pi (\mathbf{q},\omega )$ contains energy resolved information on processes contributing to the electrostatic screening and is closely related to the electron energy loss function $-\Im (1/\varepsilon(q,\omega))$, which can be measured by EELS. In normal metals we have $\Im \Pi(\omega,q)\sim \omega/q$, while band insulators show $\Im \Pi(\omega,q)=0$ for energies $\omega$ less than the band gap. Undoped graphene ($\mu =0$) lies between these cases: Here, only inter-band transitions are allowed, which leads to a peak at $%
\omega =v_{\mathrm{F}}q$ in the spectrum of $\mathrm{Im}~\Pi (\mathbf{q}%
,\omega )$. Below the energy $\omega =v_{\mathrm{F}}q$ electron-hole excitations are forbidden, as the energy exchange has to permit the 
momentum exchange $\hbar q$. In the presence of hydrogen impurities, the midgap impurity band raises the possibility of electron-hole excitations
in this forbidden region with spectral weight proportional to the impurity concentration $n_{i}$ (see Fig. \ref%
{dpim_miu}(a)). The peak at $\omega =v_{\mathrm{F}}q$ is blue
shifted and smeared out with larger impurity concentration. 

For fixed concentration of impurities, the slope of the low-frequency dependence of $\Im \Pi(\omega,q)$ increases with decreasing $q$ for $q>q_c$ (see Fig. \ref{dpim_miu}(b)) but remains constant with for $q<q_c$. We have 
$\Im \Pi(\omega,q)\sim \omega/q$ for $q>q_c$ and $\Im \Pi(\omega,q)\sim \omega/q_c$ for $q<q_c$. Thus, the dynamical polarization function also shows bad metal characteristics with electronic-excitations being available at arbitrarily low energies but with non-metallic behavior at long wavelength/small wave vectors
$q < q_c$. 
This bad metal behavior of chemically functionalized graphene is detectable in EELS experiments, as Fig. \ref{dpim_miu} (c-d) show. The impurity band leads to an electron loss signal $-\Im(1/\varepsilon(q,\omega))>0$ in the region $\omega<v_F q$ which is energetically forbidden in pristine graphene. This loss signal increases with impurity concentration (c.f. \ref{dpim_miu} (c)). As for $\Im \Pi(\omega,q)$, we find metallic $\omega$-dependence, i.e. $-\Im(1/\varepsilon(q,\omega))\sim\omega$ for $\omega\to 0$ (c.f. \ref{dpim_miu} (d)). However, the momentum transfer dependence of the EELS signal is clearly non-metallic: we find $-\Im(1/\varepsilon(q,\omega))$ increasing for $q\to 0$, whereas $-\Im(1/\varepsilon(q,\omega))$ is expected to approach a constant for $q\to 0$ in the case of a normal two-dimensional metal \footnote{A normal metal yields $\Re \Pi(q\to 0,\omega=0)=\kappa$ and $\Im \Pi(q\to 0,\omega\to 0)=\xi\omega/q$ with constants $\kappa$ and $\xi$. In RPA, we have $\varepsilon(q,\omega)=1-(2\pi e^2/q)\Pi(q,\omega)$ and consequently $\Re \varepsilon(q,\omega)\gg \Im \varepsilon(q,\omega)$ if $\kappa q\gg \xi\omega$. At sufficiently small frequencies it is thus $-\Im 1/\varepsilon(q,\omega)=\Im \varepsilon(q,\omega)/(\Re \varepsilon(q,\omega))^2\sim \omega q^0$.}.

In conclusion, we have studied electronic screening in graphene with resonant impurities. We show that the quasilocalized states in the vicinity of
neutrality point lead to \textquotedblleft bad metal\textquotedblright\ behavior. Resonant impurities make screening more efficient and render graphene metallic but only up to a certain critical length scale $l_c$. This scale is determined by Anderson localization and corresponds to the inter-impurity spacing. As a consequence, scattering of the graphene electrons by external potentials, which vary on length scales which are large compared to the interatomic spacing but smaller than $l_c$, is suppressed due to screening by the midgap states.

Support by the Stichting Fundamenteel Onderzoek der Materie (FOM), the Netherlands National Computing Facilities foundation (NCF) and by DFG through SPP 1459 are
acknowledged. TOW thanks S. Kettemann for useful discussions and RU Nijmegen for hospitality during the visit, when the ideas presented in this work were conceived.

\bibliographystyle{apsrev4-1}
\bibliography{BibliogrGrafeno}

\newcommand{\npb}{Nucl. Phys.}\newcommand{\adv}{Adv.
  Phys.}\newcommand{\epl}{Europhys. Lett.}
\begin{thebibliography}{29}%
\makeatletter
\providecommand \@ifxundefined [1]{%
 \@ifx{#1\undefined}
}%
\providecommand \@ifnum [1]{%
 \ifnum #1\expandafter \@firstoftwo
 \else \expandafter \@secondoftwo
 \fi
}%
\providecommand \@ifx [1]{%
 \ifx #1\expandafter \@firstoftwo
 \else \expandafter \@secondoftwo
 \fi
}%
\providecommand \natexlab [1]{#1}%
\providecommand \enquote  [1]{``#1''}%
\providecommand \bibnamefont  [1]{#1}%
\providecommand \bibfnamefont [1]{#1}%
\providecommand \citenamefont [1]{#1}%
\providecommand \href@noop [0]{\@secondoftwo}%
\providecommand \href [0]{\begingroup \@sanitize@url \@href}%
\providecommand \@href[1]{\@@startlink{#1}\@@href}%
\providecommand \@@href[1]{\endgroup#1\@@endlink}%
\providecommand \@sanitize@url [0]{\catcode `\\12\catcode `\$12\catcode
  `\&12\catcode `\#12\catcode `\^12\catcode `\_12\catcode `\%12\relax}%
\providecommand \@@startlink[1]{}%
\providecommand \@@endlink[0]{}%
\providecommand \url  [0]{\begingroup\@sanitize@url \@url }%
\providecommand \@url [1]{\endgroup\@href {#1}{\urlprefix }}%
\providecommand \urlprefix  [0]{URL }%
\providecommand \Eprint [0]{\href }%
\providecommand \doibase [0]{http://dx.doi.org/}%
\providecommand \selectlanguage [0]{\@gobble}%
\providecommand \bibinfo  [0]{\@secondoftwo}%
\providecommand \bibfield  [0]{\@secondoftwo}%
\providecommand \translation [1]{[#1]}%
\providecommand \BibitemOpen [0]{}%
\providecommand \bibitemStop [0]{}%
\providecommand \bibitemNoStop [0]{.\EOS\space}%
\providecommand \EOS [0]{\spacefactor3000\relax}%
\providecommand \BibitemShut  [1]{\csname bibitem#1\endcsname}%
\let\auto@bib@innerbib\@empty
\bibitem [{\citenamefont {Sarma}\ \emph {et~al.}(2011)\citenamefont {Sarma},
  \citenamefont {Adam}, \citenamefont {Huang},\ and\ \citenamefont
  {Rossi}}]{DasSarma_RMP11}%
  \BibitemOpen
  \bibfield  {author} {\bibinfo {author} {\bibfnamefont {S.~D.}\ \bibnamefont
  {Sarma}}, \bibinfo {author} {\bibfnamefont {S.}~\bibnamefont {Adam}},
  \bibinfo {author} {\bibfnamefont {E.~H.}\ \bibnamefont {Huang}}, \ and\
  \bibinfo {author} {\bibfnamefont {E.}~\bibnamefont {Rossi}},\ }\href@noop {}
  {\bibfield  {journal} {\bibinfo  {journal} {\rmp}\ }\textbf {\bibinfo
  {volume} {83}},\ \bibinfo {pages} {407} (\bibinfo {year} {2011})}\BibitemShut
  {NoStop}%
\bibitem [{\citenamefont {Katsnelson}(2012)}]{Katsnelson_book}%
  \BibitemOpen
  \bibfield  {author} {\bibinfo {author} {\bibfnamefont {M.~I.}\ \bibnamefont
  {Katsnelson}},\ }\href@noop {} {\emph {\bibinfo {title} {Graphene: Carbon in
  Two Dimensions}}}\ (\bibinfo  {publisher} {Cambridge Univ. Press,
  Cambridge},\ \bibinfo {year} {2012})\BibitemShut {NoStop}%
\bibitem [{\citenamefont {Gonzalez}\ \emph {et~al.}(1994)\citenamefont
  {Gonzalez}, \citenamefont {Guinea},\ and\ \citenamefont
  {Vozmediano}}]{Guinea_94}%
  \BibitemOpen
  \bibfield  {author} {\bibinfo {author} {\bibfnamefont {J.}~\bibnamefont
  {Gonzalez}}, \bibinfo {author} {\bibfnamefont {F.}~\bibnamefont {Guinea}}, \
  and\ \bibinfo {author} {\bibfnamefont {M.}~\bibnamefont {Vozmediano}},\
  }\href {\doibase 10.1016/0550-3213(94)90410-3} {\bibfield  {journal}
  {\bibinfo  {journal} {Nucl. Phys. B}\ }\textbf {\bibinfo {volume} {424}},\
  \bibinfo {pages} {595 } (\bibinfo {year} {1994})}\BibitemShut {NoStop}%
\bibitem [{\citenamefont {Elias}\ \emph {et~al.}(2011)\citenamefont {Elias},
  \citenamefont {Gorbachev}, \citenamefont {Mayorov}, \citenamefont {Morozov},
  \citenamefont {Zhukov}, \citenamefont {Blake}, \citenamefont {Ponomarenko},
  \citenamefont {Grigorieva}, \citenamefont {Novoselov}, \citenamefont
  {Guinea},\ and\ \citenamefont {Geim}}]{Elias_NaturePhys10}%
  \BibitemOpen
  \bibfield  {author} {\bibinfo {author} {\bibfnamefont {D.~C.}\ \bibnamefont
  {Elias}}, \bibinfo {author} {\bibfnamefont {R.~V.}\ \bibnamefont
  {Gorbachev}}, \bibinfo {author} {\bibfnamefont {A.~S.}\ \bibnamefont
  {Mayorov}}, \bibinfo {author} {\bibfnamefont {S.~V.}\ \bibnamefont
  {Morozov}}, \bibinfo {author} {\bibfnamefont {A.~A.}\ \bibnamefont {Zhukov}},
  \bibinfo {author} {\bibfnamefont {P.}~\bibnamefont {Blake}}, \bibinfo
  {author} {\bibfnamefont {L.~A.}\ \bibnamefont {Ponomarenko}}, \bibinfo
  {author} {\bibfnamefont {I.~V.}\ \bibnamefont {Grigorieva}}, \bibinfo
  {author} {\bibfnamefont {K.~S.}\ \bibnamefont {Novoselov}}, \bibinfo {author}
  {\bibfnamefont {F.}~\bibnamefont {Guinea}}, \ and\ \bibinfo {author}
  {\bibfnamefont {A.~K.}\ \bibnamefont {Geim}},\ }\href@noop {} {\bibfield
  {journal} {\bibinfo  {journal} {Nature Phys.}\ }\textbf {\bibinfo {volume}
  {7}},\ \bibinfo {pages} {701} (\bibinfo {year} {2011})}\BibitemShut {NoStop}%
\bibitem [{\citenamefont {Drut}\ and\ \citenamefont
  {L\"ahde}(2009)}]{Drut_Lahnde_PRL2009}%
  \BibitemOpen
  \bibfield  {author} {\bibinfo {author} {\bibfnamefont {J.~E.}\ \bibnamefont
  {Drut}}\ and\ \bibinfo {author} {\bibfnamefont {T.~A.}\ \bibnamefont
  {L\"ahde}},\ }\href {\doibase 10.1103/PhysRevLett.102.026802} {\bibfield
  {journal} {\bibinfo  {journal} {Phys. Rev. Lett.}\ }\textbf {\bibinfo
  {volume} {102}},\ \bibinfo {pages} {026802} (\bibinfo {year}
  {2009})}\BibitemShut {NoStop}%
\bibitem [{\citenamefont {Muniz}\ \emph {et~al.}(2010)\citenamefont {Muniz},
  \citenamefont {Dahal}, \citenamefont {Balatsky},\ and\ \citenamefont
  {Haas}}]{Balatsky_Plasmons_PRB10}%
  \BibitemOpen
  \bibfield  {author} {\bibinfo {author} {\bibfnamefont {R.~A.}\ \bibnamefont
  {Muniz}}, \bibinfo {author} {\bibfnamefont {H.~P.}\ \bibnamefont {Dahal}},
  \bibinfo {author} {\bibfnamefont {A.~V.}\ \bibnamefont {Balatsky}}, \ and\
  \bibinfo {author} {\bibfnamefont {S.}~\bibnamefont {Haas}},\ }\href {\doibase
  10.1103/PhysRevB.82.081411} {\bibfield  {journal} {\bibinfo  {journal} {Phys.
  Rev. B}\ }\textbf {\bibinfo {volume} {82}},\ \bibinfo {pages} {081411}
  (\bibinfo {year} {2010})}\BibitemShut {NoStop}%
\bibitem [{\citenamefont {Pereira}\ \emph {et~al.}(2006)\citenamefont
  {Pereira}, \citenamefont {Guinea}, \citenamefont {Lopes~dos Santos},
  \citenamefont {Peres},\ and\ \citenamefont {Castro~Neto}}]{Pereira_PRL2006}%
  \BibitemOpen
  \bibfield  {author} {\bibinfo {author} {\bibfnamefont {V.~M.}\ \bibnamefont
  {Pereira}}, \bibinfo {author} {\bibfnamefont {F.}~\bibnamefont {Guinea}},
  \bibinfo {author} {\bibfnamefont {J.~M.~B.}\ \bibnamefont {Lopes~dos
  Santos}}, \bibinfo {author} {\bibfnamefont {N.~M.~R.}\ \bibnamefont {Peres}},
  \ and\ \bibinfo {author} {\bibfnamefont {A.~H.}\ \bibnamefont
  {Castro~Neto}},\ }\href {\doibase 10.1103/PhysRevLett.96.036801} {\bibfield
  {journal} {\bibinfo  {journal} {Phys. Rev. Lett.}\ }\textbf {\bibinfo
  {volume} {96}},\ \bibinfo {pages} {036801} (\bibinfo {year}
  {2006})}\BibitemShut {NoStop}%
\bibitem [{\citenamefont {Peres}\ \emph {et~al.}(2006)\citenamefont {Peres},
  \citenamefont {Guinea},\ and\ \citenamefont {Castro~Neto}}]{Peres_PRB06}%
  \BibitemOpen
  \bibfield  {author} {\bibinfo {author} {\bibfnamefont {N.~M.~R.}\
  \bibnamefont {Peres}}, \bibinfo {author} {\bibfnamefont {F.}~\bibnamefont
  {Guinea}}, \ and\ \bibinfo {author} {\bibfnamefont {A.~H.}\ \bibnamefont
  {Castro~Neto}},\ }\href {\doibase 10.1103/PhysRevB.73.125411} {\bibfield
  {journal} {\bibinfo  {journal} {Phys. Rev. B}\ }\textbf {\bibinfo {volume}
  {73}},\ \bibinfo {pages} {125411} (\bibinfo {year} {2006})}\BibitemShut
  {NoStop}%
\bibitem [{\citenamefont {Wehling}\ \emph {et~al.}(2007)\citenamefont
  {Wehling}, \citenamefont {Balatsky}, \citenamefont {Katsnelson},
  \citenamefont {Lichtenstein}, \citenamefont {Scharnberg},\ and\ \citenamefont
  {Wiesendanger}}]{Wehling_PRB07}%
  \BibitemOpen
  \bibfield  {author} {\bibinfo {author} {\bibfnamefont {T.~O.}\ \bibnamefont
  {Wehling}}, \bibinfo {author} {\bibfnamefont {A.~V.}\ \bibnamefont
  {Balatsky}}, \bibinfo {author} {\bibfnamefont {M.~I.}\ \bibnamefont
  {Katsnelson}}, \bibinfo {author} {\bibfnamefont {A.~I.}\ \bibnamefont
  {Lichtenstein}}, \bibinfo {author} {\bibfnamefont {K.}~\bibnamefont
  {Scharnberg}}, \ and\ \bibinfo {author} {\bibfnamefont {R.}~\bibnamefont
  {Wiesendanger}},\ }\href@noop {} {\bibfield  {journal} {\bibinfo  {journal}
  {Phys. Rev. B}\ }\textbf {\bibinfo {volume} {75}},\ \bibinfo {pages} {125425}
  (\bibinfo {year} {2007})}\BibitemShut {NoStop}%
\bibitem [{\citenamefont {Wehling}\ \emph {et~al.}(2009)\citenamefont
  {Wehling}, \citenamefont {Katsnelson},\ and\ \citenamefont
  {Lichtenstein}}]{Wehling_CPL09}%
  \BibitemOpen
  \bibfield  {author} {\bibinfo {author} {\bibfnamefont {T.~O.}\ \bibnamefont
  {Wehling}}, \bibinfo {author} {\bibfnamefont {M.~I.}\ \bibnamefont
  {Katsnelson}}, \ and\ \bibinfo {author} {\bibfnamefont {A.~I.}\ \bibnamefont
  {Lichtenstein}},\ }\href@noop {} {\bibfield  {journal} {\bibinfo  {journal}
  {Chem. Phys. Lett.}\ }\textbf {\bibinfo {volume} {476}},\ \bibinfo {pages}
  {125} (\bibinfo {year} {2009})}\BibitemShut {NoStop}%
\bibitem [{\citenamefont {Wehling}\ \emph {et~al.}(2010)\citenamefont
  {Wehling}, \citenamefont {Yuan}, \citenamefont {Lichtenstein}, \citenamefont
  {Geim},\ and\ \citenamefont {Katsnelson}}]{WK10}%
  \BibitemOpen
  \bibfield  {author} {\bibinfo {author} {\bibfnamefont {T.~O.}\ \bibnamefont
  {Wehling}}, \bibinfo {author} {\bibfnamefont {S.}~\bibnamefont {Yuan}},
  \bibinfo {author} {\bibfnamefont {A.~I.}\ \bibnamefont {Lichtenstein}},
  \bibinfo {author} {\bibfnamefont {A.~K.}\ \bibnamefont {Geim}}, \ and\
  \bibinfo {author} {\bibfnamefont {M.~I.}\ \bibnamefont {Katsnelson}},\
  }\href@noop {} {\bibfield  {journal} {\bibinfo  {journal} {Phys. Rev. Lett.}\
  }\textbf {\bibinfo {volume} {105}},\ \bibinfo {pages} {056802} (\bibinfo
  {year} {2010})}\BibitemShut {NoStop}%
\bibitem [{\citenamefont {Elias}\ \emph {et~al.}(2009)\citenamefont {Elias},
  \citenamefont {Nair}, \citenamefont {Mohiuddin}, \citenamefont {Morozov},
  \citenamefont {Blake}, \citenamefont {Halsall}, \citenamefont {Ferrari},
  \citenamefont {Boukhvalov}, \citenamefont {Katsnelson}, \citenamefont
  {Geim},\ and\ \citenamefont {Novoselov}}]{Elias_Science09}%
  \BibitemOpen
  \bibfield  {author} {\bibinfo {author} {\bibfnamefont {D.~C.}\ \bibnamefont
  {Elias}}, \bibinfo {author} {\bibfnamefont {R.~R.}\ \bibnamefont {Nair}},
  \bibinfo {author} {\bibfnamefont {T.~M.~G.}\ \bibnamefont {Mohiuddin}},
  \bibinfo {author} {\bibfnamefont {S.~V.}\ \bibnamefont {Morozov}}, \bibinfo
  {author} {\bibfnamefont {P.}~\bibnamefont {Blake}}, \bibinfo {author}
  {\bibfnamefont {M.~P.}\ \bibnamefont {Halsall}}, \bibinfo {author}
  {\bibfnamefont {A.~C.}\ \bibnamefont {Ferrari}}, \bibinfo {author}
  {\bibfnamefont {D.~W.}\ \bibnamefont {Boukhvalov}}, \bibinfo {author}
  {\bibfnamefont {M.~I.}\ \bibnamefont {Katsnelson}}, \bibinfo {author}
  {\bibfnamefont {A.~K.}\ \bibnamefont {Geim}}, \ and\ \bibinfo {author}
  {\bibfnamefont {K.~S.}\ \bibnamefont {Novoselov}},\ }\href {\doibase
  10.1126/science.1167130} {\bibfield  {journal} {\bibinfo  {journal}
  {Science}\ }\textbf {\bibinfo {volume} {323}},\ \bibinfo {pages} {610}
  (\bibinfo {year} {2009})}\BibitemShut {NoStop}%
\bibitem [{\citenamefont {Nair}\ \emph {et~al.}(2010)\citenamefont {Nair},
  \citenamefont {Ren}, \citenamefont {Jalil}, \citenamefont {Riaz},
  \citenamefont {Kravets}, \citenamefont {Britnell}, \citenamefont {Blake},
  \citenamefont {Schedin}, \citenamefont {Mayorov}, \citenamefont {Yuan},
  \citenamefont {Katsnelson}, \citenamefont {Cheng}, \citenamefont
  {Strupinski}, \citenamefont {Bulusheva}, \citenamefont {Okotrub},
  \citenamefont {Grigorieva}, \citenamefont {Grigorenko}, \citenamefont
  {Novoselov},\ and\ \citenamefont {Geim}}]{Nair_small2010}%
  \BibitemOpen
  \bibfield  {author} {\bibinfo {author} {\bibfnamefont {R.~R.}\ \bibnamefont
  {Nair}}, \bibinfo {author} {\bibfnamefont {W.}~\bibnamefont {Ren}}, \bibinfo
  {author} {\bibfnamefont {R.}~\bibnamefont {Jalil}}, \bibinfo {author}
  {\bibfnamefont {I.}~\bibnamefont {Riaz}}, \bibinfo {author} {\bibfnamefont
  {V.~G.}\ \bibnamefont {Kravets}}, \bibinfo {author} {\bibfnamefont
  {L.}~\bibnamefont {Britnell}}, \bibinfo {author} {\bibfnamefont
  {P.}~\bibnamefont {Blake}}, \bibinfo {author} {\bibfnamefont
  {F.}~\bibnamefont {Schedin}}, \bibinfo {author} {\bibfnamefont {A.~S.}\
  \bibnamefont {Mayorov}}, \bibinfo {author} {\bibfnamefont {S.}~\bibnamefont
  {Yuan}}, \bibinfo {author} {\bibfnamefont {M.~I.}\ \bibnamefont
  {Katsnelson}}, \bibinfo {author} {\bibfnamefont {H.-M.}\ \bibnamefont
  {Cheng}}, \bibinfo {author} {\bibfnamefont {W.}~\bibnamefont {Strupinski}},
  \bibinfo {author} {\bibfnamefont {L.~G.}\ \bibnamefont {Bulusheva}}, \bibinfo
  {author} {\bibfnamefont {A.~V.}\ \bibnamefont {Okotrub}}, \bibinfo {author}
  {\bibfnamefont {I.~V.}\ \bibnamefont {Grigorieva}}, \bibinfo {author}
  {\bibfnamefont {A.~N.}\ \bibnamefont {Grigorenko}}, \bibinfo {author}
  {\bibfnamefont {K.~S.}\ \bibnamefont {Novoselov}}, \ and\ \bibinfo {author}
  {\bibfnamefont {A.~K.}\ \bibnamefont {Geim}},\ }\href {\doibase
  10.1002/smll.201001555} {\bibfield  {journal} {\bibinfo  {journal} {Small}\
  }\textbf {\bibinfo {volume} {6}},\ \bibinfo {pages} {2877} (\bibinfo {year}
  {2010})}\BibitemShut {NoStop}%
\bibitem [{\citenamefont {Klintenberg}\ \emph {et~al.}(2010)\citenamefont
  {Klintenberg}, \citenamefont {Lebegue}, \citenamefont {Katsnelson},\ and\
  \citenamefont {Eriksson}}]{Klintenberg_11}%
  \BibitemOpen
  \bibfield  {author} {\bibinfo {author} {\bibfnamefont {M.}~\bibnamefont
  {Klintenberg}}, \bibinfo {author} {\bibfnamefont {S.}~\bibnamefont
  {Lebegue}}, \bibinfo {author} {\bibfnamefont {M.~I.}\ \bibnamefont
  {Katsnelson}}, \ and\ \bibinfo {author} {\bibfnamefont {O.}~\bibnamefont
  {Eriksson}},\ }\href@noop {} {\bibfield  {journal} {\bibinfo  {journal}
  {Phys. Rev. B}\ }\textbf {\bibinfo {volume} {81}},\ \bibinfo {pages} {085433}
  (\bibinfo {year} {2010})}\BibitemShut {NoStop}%
\bibitem [{\citenamefont {Evers}\ and\ \citenamefont
  {Mirlin}(2008)}]{Mirlin_RMP08}%
  \BibitemOpen
  \bibfield  {author} {\bibinfo {author} {\bibfnamefont {F.}~\bibnamefont
  {Evers}}\ and\ \bibinfo {author} {\bibfnamefont {A.~D.}\ \bibnamefont
  {Mirlin}},\ }\href {\doibase 10.1103/RevModPhys.80.1355} {\bibfield
  {journal} {\bibinfo  {journal} {Rev. Mod. Phys.}\ }\textbf {\bibinfo {volume}
  {80}},\ \bibinfo {pages} {1355} (\bibinfo {year} {2008})}\BibitemShut
  {NoStop}%
\bibitem [{\citenamefont {Ostrovsky}\ \emph {et~al.}(2010)\citenamefont
  {Ostrovsky}, \citenamefont {Titov}, \citenamefont {Bera}, \citenamefont
  {Gornyi},\ and\ \citenamefont {Mirlin}}]{Ostrovsky_PRL2010}%
  \BibitemOpen
  \bibfield  {author} {\bibinfo {author} {\bibfnamefont {P.~M.}\ \bibnamefont
  {Ostrovsky}}, \bibinfo {author} {\bibfnamefont {M.}~\bibnamefont {Titov}},
  \bibinfo {author} {\bibfnamefont {S.}~\bibnamefont {Bera}}, \bibinfo {author}
  {\bibfnamefont {I.~V.}\ \bibnamefont {Gornyi}}, \ and\ \bibinfo {author}
  {\bibfnamefont {A.~D.}\ \bibnamefont {Mirlin}},\ }\href {\doibase
  10.1103/PhysRevLett.105.266803} {\bibfield  {journal} {\bibinfo  {journal}
  {Phys. Rev. Lett.}\ }\textbf {\bibinfo {volume} {105}},\ \bibinfo {pages}
  {266803} (\bibinfo {year} {2010})}\BibitemShut {NoStop}%
\bibitem [{\citenamefont {Bang}\ and\ \citenamefont
  {Chang}(2010)}]{Chang_PRB2010}%
  \BibitemOpen
  \bibfield  {author} {\bibinfo {author} {\bibfnamefont {J.}~\bibnamefont
  {Bang}}\ and\ \bibinfo {author} {\bibfnamefont {K.~J.}\ \bibnamefont
  {Chang}},\ }\href {\doibase 10.1103/PhysRevB.81.193412} {\bibfield  {journal}
  {\bibinfo  {journal} {Phys. Rev. B}\ }\textbf {\bibinfo {volume} {81}},\
  \bibinfo {pages} {193412} (\bibinfo {year} {2010})}\BibitemShut {NoStop}%
\bibitem [{\citenamefont {Yuan}\ \emph {et~al.}(2010)\citenamefont {Yuan},
  \citenamefont {De~Raedt},\ and\ \citenamefont {Katsnelson}}]{YRK10}%
  \BibitemOpen
  \bibfield  {author} {\bibinfo {author} {\bibfnamefont {S.}~\bibnamefont
  {Yuan}}, \bibinfo {author} {\bibfnamefont {H.}~\bibnamefont {De~Raedt}}, \
  and\ \bibinfo {author} {\bibfnamefont {M.~I.}\ \bibnamefont {Katsnelson}},\
  }\href {\doibase 10.1103/PhysRevB.82.115448} {\bibfield  {journal} {\bibinfo
  {journal} {Phys. Rev. B}\ }\textbf {\bibinfo {volume} {82}},\ \bibinfo
  {pages} {115448} (\bibinfo {year} {2010})}\BibitemShut {NoStop}%
\bibitem [{\citenamefont {Yuan}\ \emph {et~al.}(2011)\citenamefont {Yuan},
  \citenamefont {Rold\'an},\ and\ \citenamefont {Katsnelson}}]{YRK11}%
  \BibitemOpen
  \bibfield  {author} {\bibinfo {author} {\bibfnamefont {S.}~\bibnamefont
  {Yuan}}, \bibinfo {author} {\bibfnamefont {R.}~\bibnamefont {Rold\'an}}, \
  and\ \bibinfo {author} {\bibfnamefont {M.~I.}\ \bibnamefont {Katsnelson}},\
  }\href@noop {} {\bibfield  {journal} {\bibinfo  {journal} {Phys. Rev. B}\
  }\textbf {\bibinfo {volume} {84}},\ \bibinfo {pages} {035439} (\bibinfo
  {year} {2011})}\BibitemShut {NoStop}%
\bibitem [{\citenamefont {Pereira}\ \emph {et~al.}(2008)\citenamefont
  {Pereira}, \citenamefont {Lopes~dos Santos},\ and\ \citenamefont
  {Castro~Neto}}]{Pereira2008}%
  \BibitemOpen
  \bibfield  {author} {\bibinfo {author} {\bibfnamefont {V.~M.}\ \bibnamefont
  {Pereira}}, \bibinfo {author} {\bibfnamefont {J.~M.~B.}\ \bibnamefont
  {Lopes~dos Santos}}, \ and\ \bibinfo {author} {\bibfnamefont {A.~H.}\
  \bibnamefont {Castro~Neto}},\ }\href {\doibase 10.1103/PhysRevB.77.115109}
  {\bibfield  {journal} {\bibinfo  {journal} {Phys. Rev. B}\ }\textbf {\bibinfo
  {volume} {77}},\ \bibinfo {pages} {115109} (\bibinfo {year}
  {2008})}\BibitemShut {NoStop}%
\bibitem [{\citenamefont {Ugeda}\ \emph {et~al.}(2010)\citenamefont {Ugeda},
  \citenamefont {Brihuega}, \citenamefont {Guinea},\ and\ \citenamefont
  {G\'omez-Rodr\'iguez}}]{Ugeda2010}%
  \BibitemOpen
  \bibfield  {author} {\bibinfo {author} {\bibfnamefont {M.~M.}\ \bibnamefont
  {Ugeda}}, \bibinfo {author} {\bibfnamefont {I.}~\bibnamefont {Brihuega}},
  \bibinfo {author} {\bibfnamefont {F.}~\bibnamefont {Guinea}}, \ and\ \bibinfo
  {author} {\bibfnamefont {J.~M.}\ \bibnamefont {G\'omez-Rodr\'iguez}},\ }\href
  {\doibase 10.1103/PhysRevLett.104.096804} {\bibfield  {journal} {\bibinfo
  {journal} {Phys. Rev. Lett.}\ }\textbf {\bibinfo {volume} {104}},\ \bibinfo
  {pages} {096804} (\bibinfo {year} {2010})}\BibitemShut {NoStop}%
\bibitem [{\citenamefont {Haberer}\ \emph {et~al.}(2011)\citenamefont
  {Haberer}, \citenamefont {Petaccia}, \citenamefont {Farjam}, \citenamefont
  {Taioli}, \citenamefont {Jafari}, \citenamefont {Nefedov}, \citenamefont
  {Zhang}, \citenamefont {Calliari}, \citenamefont {Scarduelli}, \citenamefont
  {Dora}, \citenamefont {Vyalikh}, \citenamefont {Pichler}, \citenamefont
  {W\"oll}, \citenamefont {Alf\`e}, \citenamefont {Simonucci}, \citenamefont
  {Dresselhaus}, \citenamefont {Knupfer}, \citenamefont {B\"uchner},\ and\
  \citenamefont {Gr\"uneis}}]{Haberer2011}%
  \BibitemOpen
  \bibfield  {author} {\bibinfo {author} {\bibfnamefont {D.}~\bibnamefont
  {Haberer}}, \bibinfo {author} {\bibfnamefont {L.}~\bibnamefont {Petaccia}},
  \bibinfo {author} {\bibfnamefont {M.}~\bibnamefont {Farjam}}, \bibinfo
  {author} {\bibfnamefont {S.}~\bibnamefont {Taioli}}, \bibinfo {author}
  {\bibfnamefont {S.~A.}\ \bibnamefont {Jafari}}, \bibinfo {author}
  {\bibfnamefont {A.}~\bibnamefont {Nefedov}}, \bibinfo {author} {\bibfnamefont
  {W.}~\bibnamefont {Zhang}}, \bibinfo {author} {\bibfnamefont
  {L.}~\bibnamefont {Calliari}}, \bibinfo {author} {\bibfnamefont
  {G.}~\bibnamefont {Scarduelli}}, \bibinfo {author} {\bibfnamefont
  {B.}~\bibnamefont {Dora}}, \bibinfo {author} {\bibfnamefont {D.~V.}\
  \bibnamefont {Vyalikh}}, \bibinfo {author} {\bibfnamefont {T.}~\bibnamefont
  {Pichler}}, \bibinfo {author} {\bibfnamefont {C.}~\bibnamefont {W\"oll}},
  \bibinfo {author} {\bibfnamefont {D.}~\bibnamefont {Alf\`e}}, \bibinfo
  {author} {\bibfnamefont {S.}~\bibnamefont {Simonucci}}, \bibinfo {author}
  {\bibfnamefont {M.~S.}\ \bibnamefont {Dresselhaus}}, \bibinfo {author}
  {\bibfnamefont {M.}~\bibnamefont {Knupfer}}, \bibinfo {author} {\bibfnamefont
  {B.}~\bibnamefont {B\"uchner}}, \ and\ \bibinfo {author} {\bibfnamefont
  {A.}~\bibnamefont {Gr\"uneis}},\ }\href {\doibase 10.1103/PhysRevB.83.165433}
  {\bibfield  {journal} {\bibinfo  {journal} {Phys. Rev. B}\ }\textbf {\bibinfo
  {volume} {83}},\ \bibinfo {pages} {165433} (\bibinfo {year}
  {2011})}\BibitemShut {NoStop}%
\bibitem [{\citenamefont {Kubo}(1957)}]{K57}%
  \BibitemOpen
  \bibfield  {author} {\bibinfo {author} {\bibfnamefont {R.}~\bibnamefont
  {Kubo}},\ }\href {\doibase 10.1143/JPSJ.12.570} {\bibfield  {journal}
  {\bibinfo  {journal} {J. Phys. Soc. Jpn.}\ }\textbf {\bibinfo {volume}
  {12}},\ \bibinfo {pages} {570} (\bibinfo {year} {1957})}\BibitemShut
  {NoStop}%
\bibitem [{\citenamefont {Wunsch}\ \emph {et~al.}(2006)\citenamefont {Wunsch},
  \citenamefont {Stauber}, \citenamefont {Sols},\ and\ \citenamefont
  {Guinea}}]{WSSG06}%
  \BibitemOpen
  \bibfield  {author} {\bibinfo {author} {\bibfnamefont {B.}~\bibnamefont
  {Wunsch}}, \bibinfo {author} {\bibfnamefont {T.}~\bibnamefont {Stauber}},
  \bibinfo {author} {\bibfnamefont {F.}~\bibnamefont {Sols}}, \ and\ \bibinfo
  {author} {\bibfnamefont {F.}~\bibnamefont {Guinea}},\ }\href@noop {}
  {\bibfield  {journal} {\bibinfo  {journal} {New Journal of Physics}\ }\textbf
  {\bibinfo {volume} {8}},\ \bibinfo {pages} {318} (\bibinfo {year}
  {2006})}\BibitemShut {NoStop}%
\bibitem [{Note1()}]{Note1}%
  \BibitemOpen
  \bibinfo {note} {Due to limitations of present computer power, it is
  numerically too expensive to get reliable values of the polarization function
  for wave vectors $q<0.05/a$}\BibitemShut {NoStop}%
\bibitem [{\citenamefont {Shklovskii}\ and\ \citenamefont
  {Efros}(1984)}]{Shklovskii_Efros}%
  \BibitemOpen
  \bibfield  {author} {\bibinfo {author} {\bibfnamefont {B.~I.}\ \bibnamefont
  {Shklovskii}}\ and\ \bibinfo {author} {\bibfnamefont {A.~L.}\ \bibnamefont
  {Efros}},\ }\href@noop {} {\emph {\bibinfo {title} {Electronic Properties of
  Doped Semiconductors}}}\ (\bibinfo  {publisher} {Springer, Berlin},\ \bibinfo
  {year} {1984})\BibitemShut {NoStop}%
\bibitem [{\citenamefont {Lee}\ and\ \citenamefont
  {Ramakrishnan}(1985)}]{Lee_RMP85}%
  \BibitemOpen
  \bibfield  {author} {\bibinfo {author} {\bibfnamefont {P.~A.}\ \bibnamefont
  {Lee}}\ and\ \bibinfo {author} {\bibfnamefont {T.~V.}\ \bibnamefont
  {Ramakrishnan}},\ }\href {\doibase 10.1103/RevModPhys.57.287} {\bibfield
  {journal} {\bibinfo  {journal} {Rev. Mod. Phys.}\ }\textbf {\bibinfo {volume}
  {57}},\ \bibinfo {pages} {287} (\bibinfo {year} {1985})}\BibitemShut
  {NoStop}%
\bibitem [{\citenamefont {Lherbier}\ \emph {et~al.}(2011)\citenamefont
  {Lherbier}, \citenamefont {Dubois}, \citenamefont {Declerck}, \citenamefont
  {Roche}, \citenamefont {Niquet},\ and\ \citenamefont
  {Charlier}}]{Roche_PRL11}%
  \BibitemOpen
  \bibfield  {author} {\bibinfo {author} {\bibfnamefont {A.}~\bibnamefont
  {Lherbier}}, \bibinfo {author} {\bibfnamefont {S.~M.-M.}\ \bibnamefont
  {Dubois}}, \bibinfo {author} {\bibfnamefont {X.}~\bibnamefont {Declerck}},
  \bibinfo {author} {\bibfnamefont {S.}~\bibnamefont {Roche}}, \bibinfo
  {author} {\bibfnamefont {Y.-M.}\ \bibnamefont {Niquet}}, \ and\ \bibinfo
  {author} {\bibfnamefont {J.-C.}\ \bibnamefont {Charlier}},\ }\href {\doibase
  10.1103/PhysRevLett.106.046803} {\bibfield  {journal} {\bibinfo  {journal}
  {Phys. Rev. Lett.}\ }\textbf {\bibinfo {volume} {106}},\ \bibinfo {pages}
  {046803} (\bibinfo {year} {2011})}\BibitemShut {NoStop}%
\bibitem [{Note2()}]{Note2}%
  \BibitemOpen
  \bibinfo {note} {A normal metal yields ${\protect \rm Re}\protect \tmspace
  +\thinmuskip {.1667em}\Pi (q\to 0,\omega =0)=\kappa $ and ${\protect \rm
  Im}\protect \tmspace +\thinmuskip {.1667em}\Pi (q\to 0,\omega \to 0)=\xi
  \omega /q$ with constants $\kappa $ and $\xi $. In RPA, we have $\varepsilon
  (q,\omega )=1-(2\pi e^2/q)\Pi (q,\omega )$ and consequently ${\protect \rm
  Re}\protect \tmspace +\thinmuskip {.1667em}\varepsilon (q,\omega )\gg
  {\protect \rm Im}\protect \tmspace +\thinmuskip {.1667em}\varepsilon
  (q,\omega )$ if $\kappa q\gg \xi \omega $. At sufficiently small frequencies
  it is thus $-{\protect \rm Im}\protect \tmspace +\thinmuskip
  {.1667em}1/\varepsilon (q,\omega )={\protect \rm Im}\protect \tmspace
  +\thinmuskip {.1667em}\varepsilon (q,\omega )/({\protect \rm Re}\protect
  \tmspace +\thinmuskip {.1667em}\varepsilon (q,\omega ))^2\sim \omega
  q^0$.}\BibitemShut {Stop}%
\end{thebibliography}%

\end{document}